\documentclass[reprint,superscriptaddress,aps,prl,showpacs]{revtex4-1}
\usepackage{graphicx}

\begin{document}

\title{Metal frame as local protection of superconducting films from thermomagnetic avalanches }

\author{P.~ Mikheenko}
\affiliation{Department of Physics,
University of Oslo,  P.O. Box
1048 Blindern, 0316 Oslo, Norway}

\author{J. I. Vestg{\aa}rden}
\affiliation{Department of Physics, University of Oslo, P.O. Box
1048 Blindern, 0316 Oslo, Norway}

\author{S. Chaudhuri}
\affiliation{Nanoscience Center, Department of Physics, P.O. Box
35, University of Jyv\"{a}askyl\"{a}, FIN-40014
Jyv\"{a}askyl\"{a}, Finland}

\author{I. J. Maasilta}
\affiliation{Nanoscience Center, Department of Physics, P.O. Box
35, University of Jyv\"{a}askyl\"{a}, FIN-40014
Jyv\"{a}askyl\"{a}, Finland}

\author{Y. M. Galperin}
\affiliation{Department of Physics, University
of Oslo, P.O. Box
1048 Blindern, 0316 Oslo, Norway}
\affiliation{Physico-Technical Institute RAS, 194021 St. Petersburg,
Russian Federation}

\author{T. H. Johansen}
\affiliation{Department of Physics, University of Oslo, P.O. Box
1048 Blindern, 0316 Oslo, Norway}

\affiliation{Institute for Superconducting and Electronic
Materials, University of Wollongong, Northfields Avenue,
Wollongong, NSW 2522, Australia}

\begin{abstract}
Thermomagnetic avalanches in superconducting films propagating extremely fast while forming
unpredictable patterns, represent a serious threat for the performance of devices based
on such materials.
It is shown here that a normal-metal frame surrounding a selected region inside the film area
can provide efficient protection from the avalanches during their propagation stage.
Protective behavior is confirmed by magneto-optical imaging experiments on
NbN films equipped with Cu and Al frames, and also by performing numerical simulations.
Experimentally, it is found that while conventional flux creep is not affected by the
frames, the dendritic avalanches are partially or fully screened by them. The level of screening
depends on the ratio of the sheet conductance of the metal and the superconductor in the
resistive state, and for ratios much larger than unity  the screening is very efficient.

\end{abstract}

\pacs{74.25.fc, 74.25.Ha, 74.25.N-, 74.25.Op}

\maketitle

The number of applications based on superconductors increases,
and with the important role these devices
will play in technology,\cite{basicsuperc,wangy1,mikheenko11}  safe operation
 becomes crucially important. When loaded with a high electrical
current, or exposed to a strong magnetic field,   the superconductors accumulate large amounts
of energy. This energy can suddenly be released through a thermomagnetic runaway, or
avalanche, which in general is very harmful, e.g., to superconducting magnets and electronic devices.
Especially vulnerable are superconducting thin-film devices experiencing a perpendicular magnetic field,
where avalanche events may occur at fields as low as a few millitesla.\cite{johansen02, shantsev05}

The  instability that triggers such    avalanches is deeply rooted in the nature of type-II superconductors,
where magnetic flux  exists in the form of quantized vortices. If a vortex moves, it
dissipates energy causing a local temperature rise in the material. This promotes motion of the
neighboring vortices, and the positive feedback can create a massive thermomagnetic
avalanche. In thin films, magneto-optical imaging (MOI) has revealed that the instability
leads to abrupt flux motion in the form of large, often sample-spanning, dendritic structures, where each branch
propagates at a speed up to 100~km/s.\cite{leiderer93,bolz03,bolz03-2}
The phenomenon has been observed in films of many superconducting
 materials.\cite{leiderer93, duran95, johansen01, rudnev03, altshuler04-2, wimbush04, rudnev05}

Although the avalanche behavior has already undergone extensive
investigations,\cite{altshuler04} one finds from a practical viewpoint that these events
are largely out of control. Partly, this is due to (i) the unpredictability of when and
where an avalanche starts, and (ii) the unpredictable path that each branch of the avalanche will follow.
Furthermore, experimental studies of the phenomenon are severely limited by
the lack of techniques allowing to investigate the flux  dynamics on the relevant time scale
of a few nanoseconds. This is the duration of a typical event,\cite{vestgarden12-sr} and
is also the time scale when damage is done to a device.
In fact, an avalanche can permanently
damage the  superconductor, as shown in recent work on
YBa$_2$Cu$_3$O$_x$ films.\cite{baziljevich14}
Here, MOI revealed  that superconducting properties
can be lost in parts of the avalanche path, and atomic force microscopy
scans showed that the local heating can even cause complete disintegration
of the  material. Thus, from an applied perspective is it essential to find means to prevent such events from
happening, in particular, in regions of vital importance for the functionality of a device.

Previous experiments demonstrated that a uniform metal layer deposited on the
superconducting film can suppress the avalanche
activity.\cite{baziljevich02,choi05,albrecht05,treiber10,brisbois14} It was also reported that
one may selectively prevent avalanche nucleation along  the rim by coating parts of it with
a normal metal film.\cite{choi09, mikheenko15} I
 In the present work, we show that
avalanches can be stopped even after they have nucleated, i.e., during their fast and destructive
propagation stage. This is done by adding a normal metal frame surrounding a selected
internal area of the superconducting film. The efficiency of the local protection is
documented by MOI observations and numerical simulations.

\begin{figure}[t]
\centering
\includegraphics[height=7.7cm]{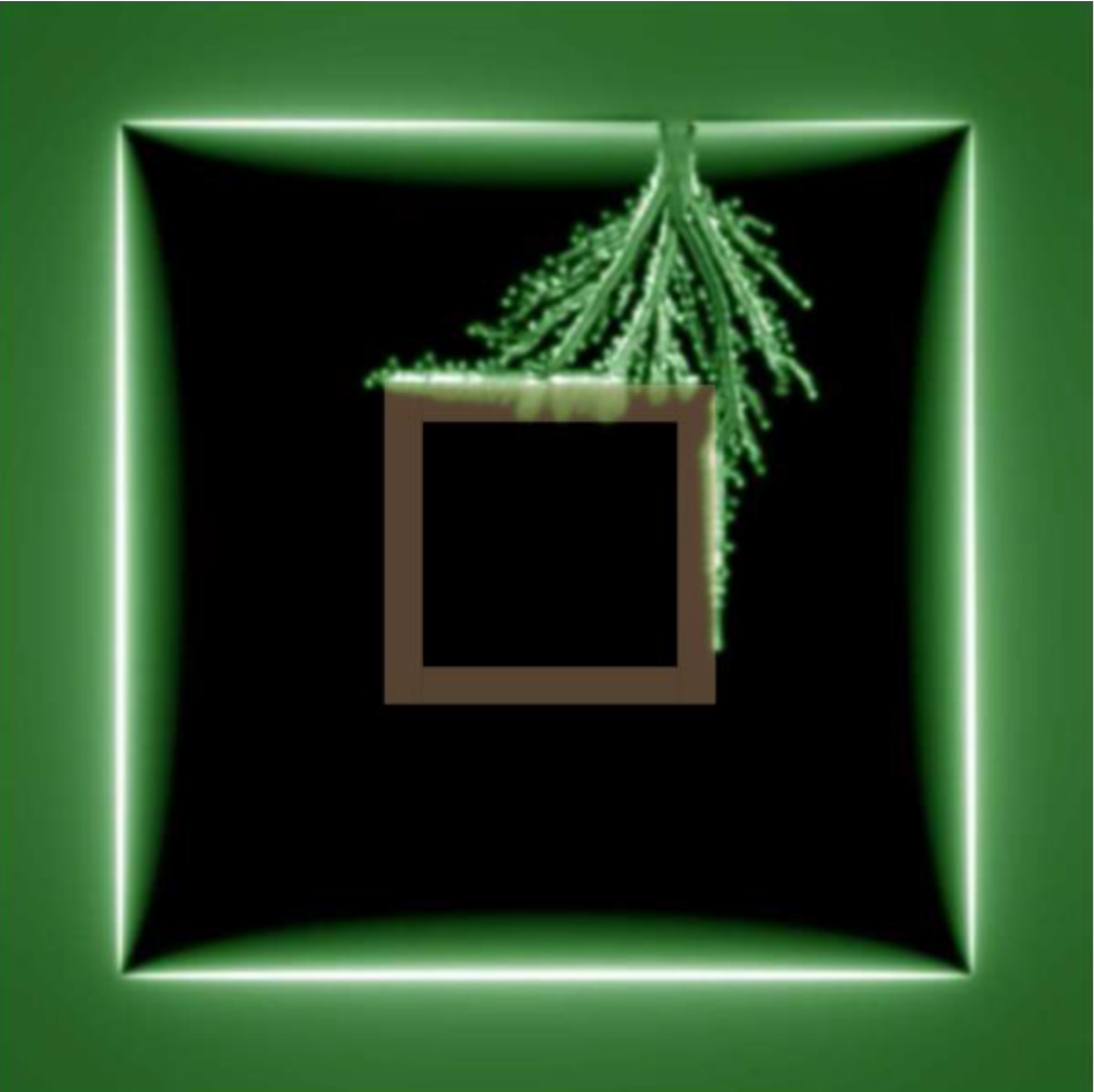}
\caption{Numerical simulation of the flux distribution in a
superconducting film, where the propagation of an avalanche
was blocked by a metal frame (light-brown).
The brightness of the green color represents the magnitude of the flux density.\label{fig1}}
\end{figure}

The sample configuration chosen in this work is a square superconducting film,
see Fig.~\ref{fig1}, where in the central part a narrow square  metal frame  (light brown color) is placed to
protect the area it surrounds.
The figure illustrates the result of a numerical simulation where such a sample is
exposed to an increasing perpendicular magnetic field.
The simulations were performed following the formalism described in
Refs.~\onlinecite{vestgarden13-sr, vestgarden13-metal},  and using typical parameters for low-resistive
(in the normal state) superconducting films.
In the figure the film edge appears as a very bright contour, showing the piling up
of the external field due to the diamagnetic response of the superconductor.
This particular image represents the flux density at a time 65~ns after an avalanche started
 from the upper film edge. Evidently, this avalanche  propagated in a branched fashion,
and the shown dendritic structure is the final flux pattern after the avalanche came to rest.
The spatially smooth penetration of flux from all the external edges shows the regular
 behavior of a square superconducting film.

From the flux distribution in Fig.~\ref{fig1} it is clear that the avalanche was
strongly influenced by the metal frame. Indeed, the frame fully prevented all the rapidly
approaching  flux branches from entering the enclosed central region. Interestingly, this
protection occurs in two different ways, namely
(i) by reflecting incoming flux branches, and (ii) by damping the flux motion taking place under the metal coating.
Note also that in the upper horizontal part of the metal frame, the flux branches hitting the frame at
approximately normal incidence are close to penetrate through the obstruction.
Also the right vertical part of the frame is activated, serving to guide some branches along the outer edge of the frame.

\begin{figure}[t]
\centering
\includegraphics[height=7.7cm]{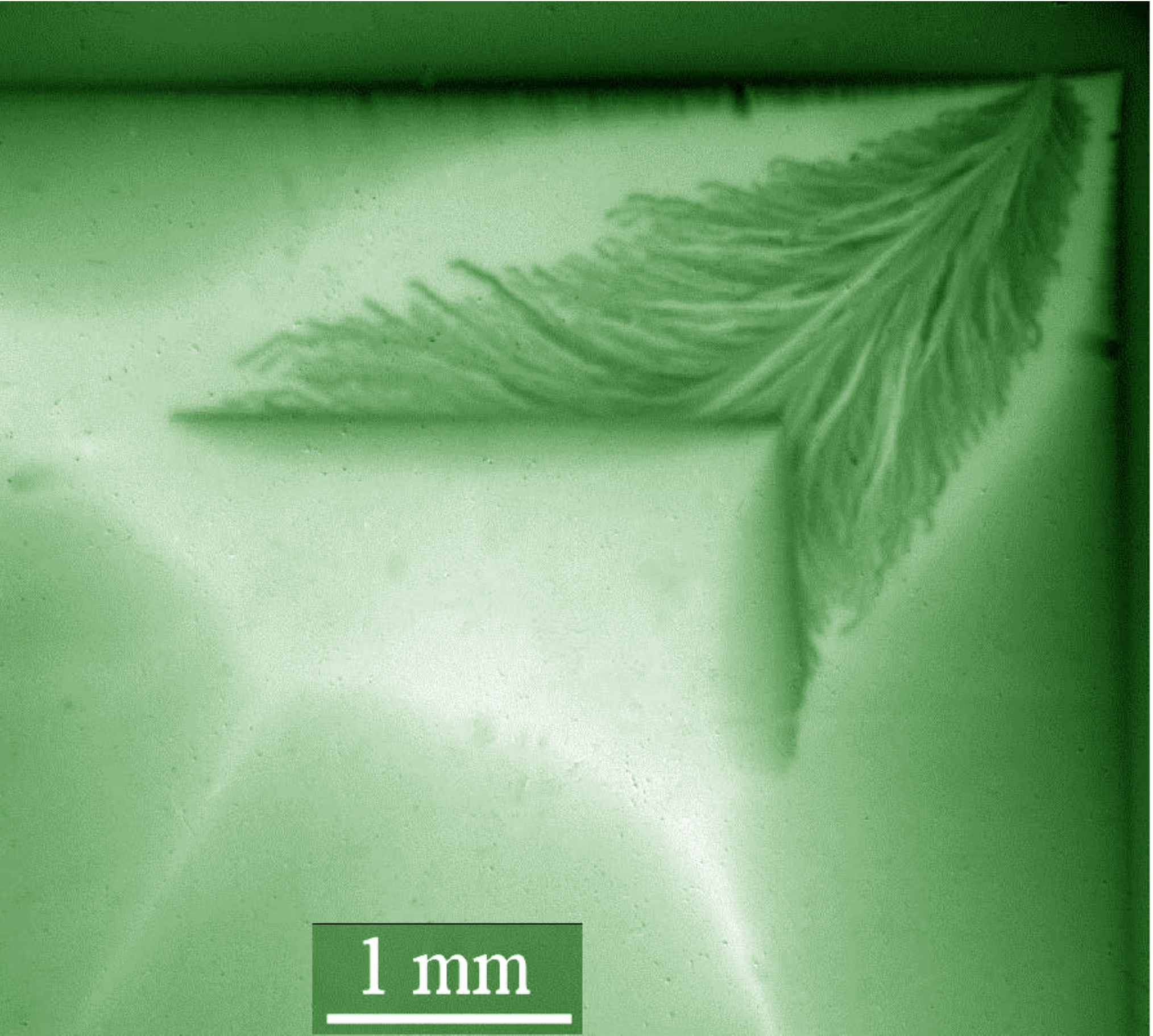}
\caption{Flux distribution in the upper right corner area of a NbN film, as the
descending field reaches 9.2~mT. The magneto-optical image was recorded using slightly
uncrossed  polarizers. \label{fig2}}
\end{figure}

To test in practice the avalanche-protection ability of a metal frame on superconducting films,
pulsed laser deposition was used to grow films of NbN on MgO (001) single crystal
substrates.\cite{chaudhuri11}
A film of thickness 170 nm  was shaped as a square with sides measuring
8~mm. A 1~$\mu$m thick Cu frame was then deposited on
 the sample, with the same relative positioning as seen in
Fig.~\ref{fig1}.  The square frame measures 4~mm externally,
and has a width of 0.5 mm.
Visualization of magnetic flux distributions accross the sample area was done by MOI
using a Faraday rotating ferrite garnet sensor plate~\cite{helseth01, helseth02} placed
directly on the sample.

Shown in Fig.~\ref{fig2} is a magneto-optical image of the flux distribution formed  by
an avalanche starting from the upper edge of an NbN film. The
sample had here first been zero-field-cooled to 4~K, and then exposed to a 12~mT
perpendicular field, which caused full penetration of flux without triggering any avalanche.
The applied field was then reduced,
and when reaching  9.2~mT the  dendritic avalanche seen in the figure occurred.

It is evident that all the branches in this avalanche were blocked by the metal
frame. Interestingly, the blocking has to a large extent the character of reflection from
the interface.\cite{mikheenko15} Another visible feature is that some of the avalanching flux, when hitting the frame,
penetrates smoothly into the metal coated area, where it finally comes to rest in a
critical-state-like distribution. The frame gives here full avalanche protection of the
enclosed area, and does it with characertistics agreeing very well with the simulation
results.

\begin{figure}[t]
\centering
\includegraphics[height=7.4cm]{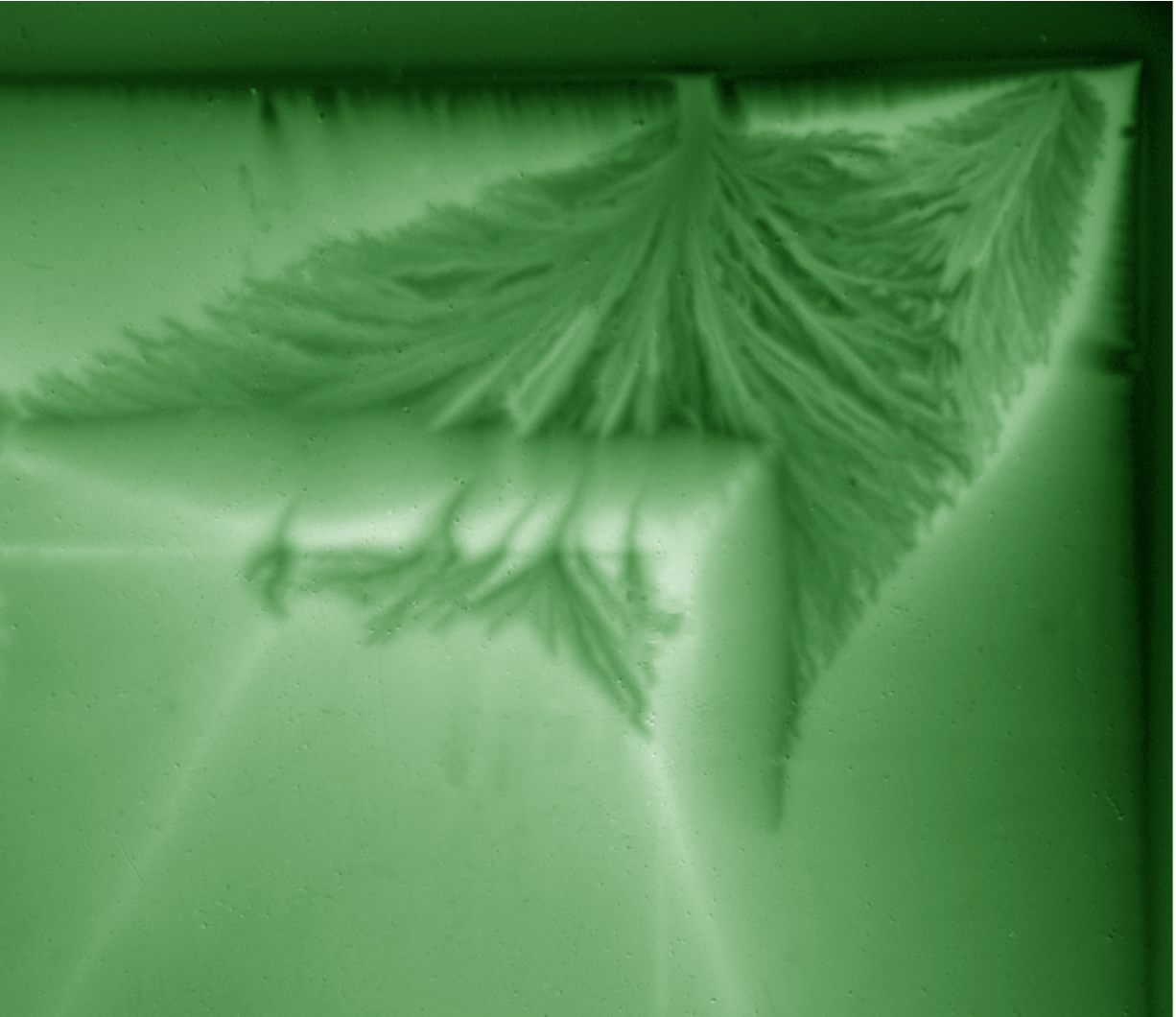}
\caption{Magnetic flux distribution as in Fig.~\ref{fig2} when the field is further
reduced to 5.8~mT. When a second, more powerful, avalanche occurs, the protection is
incomplete. \label{fig3}}
\end{figure}

When lowering the applied field further to 5.8~mT, a second avalanche occured, starting
from a different point on the upper sample edge, see Fig.~\ref{fig3}.
The new starting point caused many of the branches in the avalanche to hit the frame
at near perpendicular incidence. Evidently, some of these branches traverse the frame.
Although they leave only faint traces of their crossing,
they form again a typical branched patterns when entering the uncoated square.
Thus, in this case the metal frame was not able to provide full protection of the central area.

\begin{figure}[t]
\centering
\includegraphics[height=7.4cm]{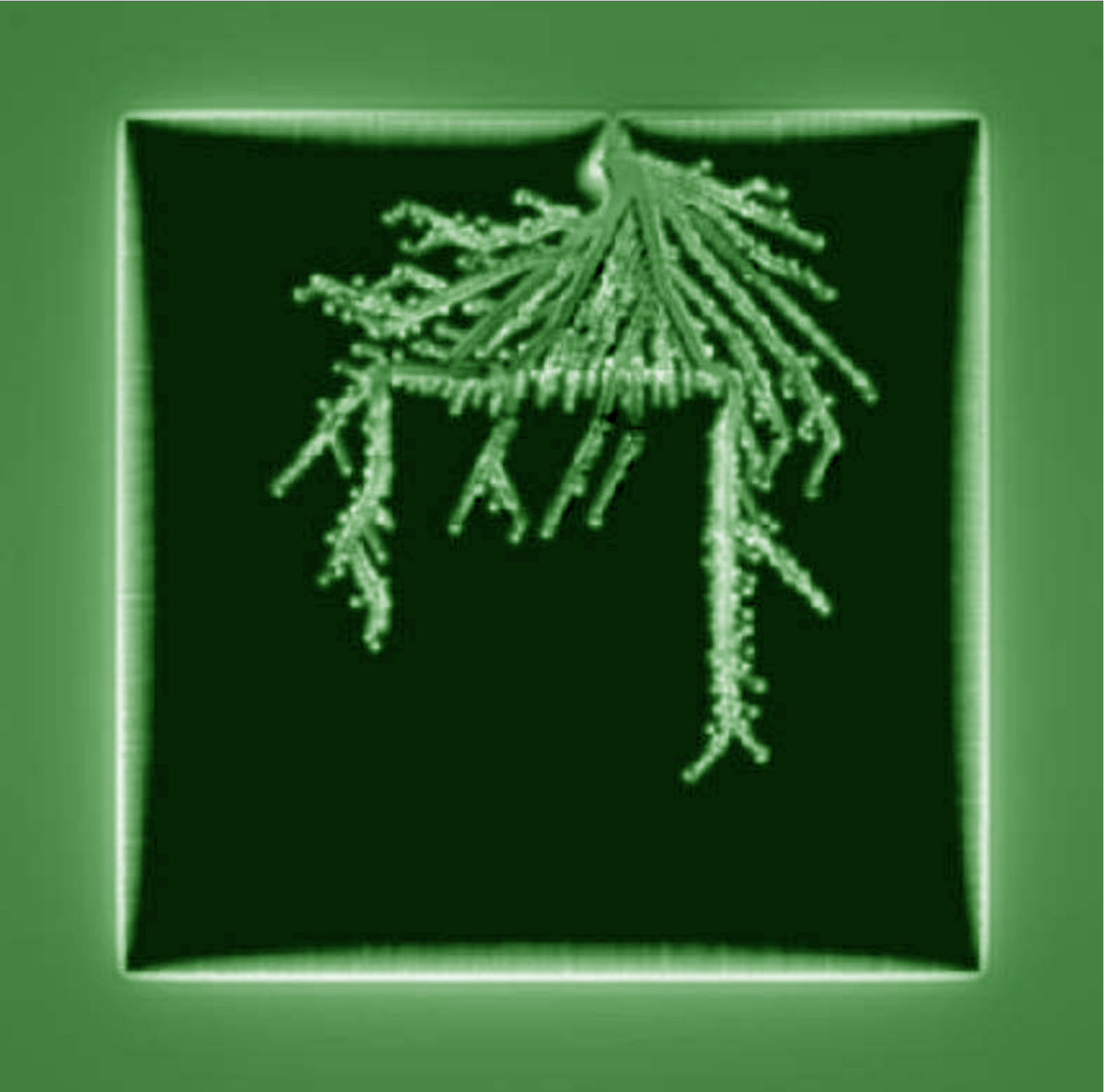}
\caption{Simulated penetration of a dendrite through the conductive frame. $S=10$.\label{fig4}}
\end{figure}

To verify the importance of the angle between the metal frame edge and the direction
of the incoming avalanche branches, simulations were performed using a sample matrix
causing the avalanche to start from a geometrical location similar to what is seen
in Fig.~\ref{fig3}. The result of the numerical calculation is shown in Fig.~\ref{fig4}.
The key features of the experimental and numerical results are again strikingly similar,
in particular the fact that several branches of the avalanche are now traversing the metal frame.
Moreover, the branches entering the area to be protected are seen to form further branching.

According to Ref.~\onlinecite{vestgarden14-sr} the electromagnetic damping of thermomagnetic
avalanches by a normal-metal coating is governed by the dimensionless parameter,
$S = (\rho_s d_m)/(\rho_m \, d_s) $.  Here, $\rho_m$ and $\rho_s$ are
resistivities of the metal and superconductor (in the resistive state), respectively, while
 $d_m$ and  $d_s$ are the thicknesses of the corresponding layers.
 In other words,  $S$ is the ratio of the normal-state sheet resistance of the superconductor, $R_s$,
and the sheet resistance of the coating metal, $R_m$.
For efficient screening the value of $S$ should be much larger than unity. In the present case,
after subtracting the contact resistance, we measured at room temperature that $R_s$
between points separated by 2.25 mm equals 7 Ohm, while $R_m$ between the same points  is 0.5 Ohm.
This gives $S \approx 14 $ for the sample displayed in Figs.~2 and 3, quite consistent
 with the fairly good avalanche protection.

An obvious way to increase the screening efficiency is to increase the thickness of the normal metal layer.
However, deposition of a layer with thickness much more than one micron is difficult since thick metal
films tend to peel off the sample. Also due to a finite skin depth in the metal at the ultra-short time
scale typical for the dendrite propagation, only part of the thickness could contribute to screening.
E.g., the skin depth in Al is $\sim 2~\mu$m for processes of duration of 1~ns,
and  $6~\mu$m for processes with a characteristic time of 10~ns.

A more reliable and flexible approach could be to use an external frame rather than one deposited on the film.
To test this concept, a frame was cut from a 13 $\mu$m thick Al technical foil, and
placed on the superconductor using the light pressure from the magneto-optical
sensor plate for stable mounting.

\begin{figure}[t]
\centering
\includegraphics[width=.8\columnwidth]{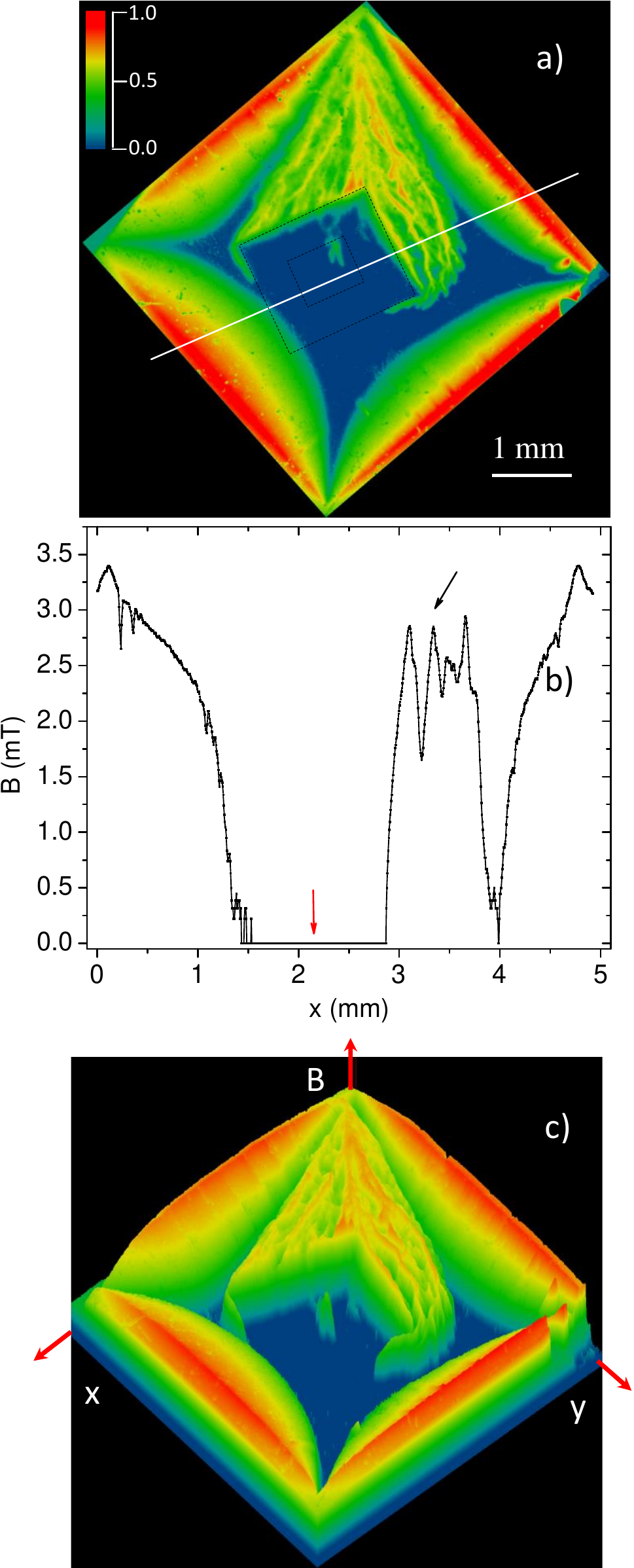}
\caption{a) A color-coded magneto-optical image of superconducting film with an
Al frame (see the dark dotted lines) in an applied magnetic
field of 2.9 mT. The numbers at the scale correspond to the image brightness in arbitrary units.
 b) The local magnetic field profile along the white line shown in a).
c) A 3D representation of the flux density in the sample. \label{fig5}}
\end{figure}

Shown in Fig.~\ref{fig5}  is the observed flux
penetration pattern inside the sample after it was initially zero-field cooled to 4~K,
and then subjected to a perpendicular field of 2.9~mT.
Panel a) displays the magneto-optical image using color coding,
where red color represents its maximum brightness.
The frame is indicated by the black dashed lines showing a slight rotation
relative to the superconductor square.

In this figure,  as in Fig.~\ref{fig3}, one sees two regimes of flux penetration -- one
consisting of avalanche  dendrites, and one showing conventional penetration forming
typical critical-state profiles.~\cite{brandt95-prl}
The avalanching flux is again strongly influenced by the frame, which prevents nearly all the
incoming branches from entering into the central uncoated area.
Instead, they disperse into a smooth flux distribution within the metal-coated part.
Only one small fragment of the flux branches reaches the central frame area.
Contrary, one sees that the flux having penetrated the sample from the lower left side of the square,
was not at all influenced by the metal frame as it entered the frame area at a much smaller velocity.

The magneto-optical image in Fig.~\ref{fig5}a was recalcu\-lated\cite{johansen96, jooss02}
into a distribution of the flux density, $B$ over the sample area.
The panel b displays the flux density along the white line drawn in panel a.
From the $B$-profile one sees that the
area enclosed by the frame has a large continuous
region of perfect screening.
 Thus, if a sensitive superconducting
device were placed there, it would not be affected by the avalanche event.
Note also that due to the finite skin depth, there is little to gain by increasing
the thickness of this frame any further.
However, improving the quality of the Al-foil and its mounting is likely to make the screening
more efficient.

Interestingly, one can see  from panel b that  the  $B$-profile across the frame  is
substantially steeper than the Bean critical-state profile in the bare sample. Therefore,
coating by a normal metal improves the screening of fast electromagnetic excitations in type-II superconductors.
Overall, with the deposition of a highly conductive normal metal layer and proper adjustment
of the device geometry, one can obtain substantial screening of dendritic avalanches,
the propagation of slowly moving magnetic flux remaining essentially unperturbed.

In conclusion, it is demonstrated that a normal metal frame added on top of a superconducting film
strongly impedes the propagation of thermomagnetic avalanches.
Using such frames, it is
possible to screen selected areas of the film from the destructive avalanche events,
while keeping unaffected all slowly moving magnetic flux, e.g., as part of the communication
with superconducting electronics located inside the area enclosed by the frame.
The experiments also reveal that the penetration of avalanche branches into the framed region
 depends on the angle of incidence. Avalanche branches hitting the frame at near normal incidence
 are much less screened than those hitting at smaller angles which are
mostly reflected.


%

\end{document}